\documentclass[12pt]{iopart}
\usepackage{graphicx}
\usepackage{epstopdf}
\usepackage[usenames]{color}
\expandafter\let\csname equation*\endcsname\relax
\expandafter\let\csname endequation*\endcsname\relax
\usepackage{amssymb,amsmath}
\usepackage[square,comma,compress,numbers]{natbib}

\newcommand{\fig}[1]{Fig.~\ref{#1}}
\newcommand{\be}[1]{\begin{equation}\label{#1}}
\newcommand{\ee}{\end{equation}}

\usepackage{float}

\begin{document}
\title{Multiple core hole formation by free-electron laser radiation in molecular nitrogen}
\author{H. I. B. Banks}
\address{Department of Physics and Astronomy, University College London, Gower Street, London WC1E 6BT, United Kingdom}
\author{D. A. Little}
\address{Department of Physics and Astronomy, University College London, Gower Street, London WC1E 6BT, United Kingdom}
\author{A. Emmanouilidou}
\address{Department of Physics and Astronomy, University College London, Gower Street, London WC1E 6BT, United Kingdom}

\begin{abstract}
We investigate the formation of multiple-core-hole states of molecular nitrogen interacting with a free-electron laser pulse. We obtain bound and continuum molecular orbitals in the single-center expansion scheme and use these orbitals to calculate photo-ionization and Auger decay rates. Using these rates, we compute the atomic ion yields generated in this interaction. We track the population of all states throughout this interaction and compute the proportion of the population which accesses different core-hole states. We also investigate the pulse parameters that favor the formation of these core-hole states for 525 eV and 1100 eV photons. 
 \end{abstract}
\date{\today}

\maketitle
 
\section{Introduction}

The production of free-electron lasers (FELs) \cite{FELhistory} with x-ray photon energies has led to new methods of investigating atoms and molecules \cite{Marangos2011,Ullrich2012}, and of imaging biomolecules \cite{Schlichting2012,Neutze,Redecke227}. The x-ray energy allows the single-photon ionization of core electrons and the creation of core-hole states. These core-hole states have lifetimes of a few femtoseconds. These states  decay via Auger processes, in which the valence electron fills in the core hole and the released energy allows another valence electron to escape. If the FEL has a high enough photon flux to ionize more than one core electron before an Auger decay takes place then multiple-core-hole states can be produced. Several studies have addressed double-core-hole (DCH) states of molecules, as they are of particular interest for chemical analysis \cite{Cederbaum1986,Tashiro2010}. The energy of these states is highly dependent on the chemical environment, making them an appropriate basis for spectroscopic measurements.
In molecules, DCH states are either single site double-core-hole states (SSDCH) with the core holes on one atomic site or two site double-core-hole states (TSDCH) with the core holes on different atomic sites. TSDCH states are particularly sensitive to their chemical environment as the repulsion and relaxation effects are dependent on the bond type \cite{Cederbaum1986}. As such, several studies have addressed these states both experimentally \cite{Berrah2011,Fang2010,Fang2012,Salen2012,Zhaunerchyk2015} and theoretically \cite{Cederbaum1986,Tashiro2010,Buth2012,Liu2016}.

Moreover, with sufficiently high intensity, three core holes can be generated before Auger decay occurs forming triple-core-hole states (TCH). Theoretical studies of TCH states include  i)  the investigation of the  formation during the interaction of atomic Argon with FEL radiation \cite{Wallis2015}; ii) the computation of the energies of TCH states for molecular nitrogen using the multi-configurational self-consistent field (MCSCF) technique \cite{Carravetta2013}; the creation of triply excited hollow states in laser-driven Lithium with subsequent autoionization to doubly excited states \cite{Madsen2001}.

In the present study, we investigate the formation of TSDCH and TCH states in FEL-driven $\mathrm{N_2}$. We use molecular bound and molecular continuum orbitals to compute single-photon ionization and Auger rates with 525 eV and 1100 eV photons. These photon energies allow us to ionize three or four core electrons, respectively. This is an advantage over previous calculations where the molecule is treated as a combination of atoms \cite{Liu2016} and over computations where atomic rather than molecular continuum orbitals are employed to describe the interaction of molecules with FEL laser pulses \cite{Inhester2016}. The use of molecular continuum orbitals versus atomic continuum orbitals will yield more accurate results for low photon-energy FEL pulses as well as short duration and high intensity FEL pulses \cite{Fang2012}. We compute atomic and molecular single-photon ionization cross sections and Auger rates and keep the nuclei fixed. We account for the fragmentation of the FEL-driven $\mathrm{N_{2}}$ through additional terms in the rate equations. In our computations, we keep track of the percentages of the populations of the final atomic ion fragments, which go through different energetically allowed sequences of intermediate molecular and atomic states; each state is determined by the electronic configuration. In what follows, we refer to these sequences as pathways. We thus register the percentage of the final atomic ion yields that transitions through DCH or TCH molecular states. Moreover, at each time step of our computations, for the DCH molecular states we project the occupied delocalized molecular orbitals onto orbitals localized on an atomic site. Thus, we determine the percentage of DCH molecular states that correspond to TSDCH and to SSDCH molecular states. Finally, we compute as a function of the intensity of the FEL pulse the percentage of the total population that accesses TSDCH, SSDCH and TCH molecular states. 







\section{Method}
\subsection{Rate equations for FEL-driven $\mathrm{N_{2}}$}

We obtain the bound molecular and bound atomic orbital wavefunctions using Molpro \cite{Molpro}. For FEL-driven $\mathrm{N_2}$ we treat the nuclei as fixed, with the distance between the nuclei set equal to the equilibrium distance, i.e. 2.08 a.u. This inter-nuclear distance is used to compute the molecular orbitals. We use the single-center expansion method \cite{Demekhin2011} to calculate the molecular continuum wavefunctions and Hartree-Fock Slater \cite{Herman1963} calculations to compute the atomic continuum wavefunctions \cite{Wallis2015}. We then use these wavefunctions to obtain single-photon ionization cross sections and Auger rates, see \cite{Banks2017}. The dissociation of the molecule is treated through terms in the rate equations. Specifically, we assume that $\mathrm{N_2^{4+}}$ and $\mathrm{N_2^{3+}}$ molecular ion states with no core holes dissociate instantaneously into $\mathrm{N^{2+}+N^{2+}}$ and $\mathrm{N^{2+}+N^{+},}$ respectively. We also assume that all states of $\mathrm{N_2^{2+}}$ dissociate with a lifetime of 100 fs \cite{Beylerian2004}. The final fragments of this latter dissociation are $\mathrm{N^{+}+N^{+}}$ and  $\mathrm{N^{2+}+N}$ with 74\% and 26\% probability, respectively. Starting from the ground state of molecular nitrogen with all orbitals fully occupied ($\mathrm{1\sigma_g^{2},1\sigma_u^{2},2\sigma_g^{2},2\sigma_u^{2},1\pi_{ux}^{2},1\pi_{uy}^{2},3\sigma_g^{2}}$), we solve the rate equations to obtain the final atomic ion yields. Moreover, by solving additional rate equations,  we compute all the possible pathways which contribute to these final atomic ion states. By pathway, we refer to the series of molecular and atomic states (electronic configurations) that are accessed before the formation of a given atomic ion yield. In what follows, we compute   the contribution of pathways that include the formation of a SSDCH, TSDCH or TCH molecular state. 

\subsection{Projection of delocalized molecular orbitals onto orbitals localized on atomic sites}\label{Projection}

As mentioned above, we use molecular orbitals to obtain  the final atomic ion yields and the populations of the pathways from the rate equations. We note that the use
of molecular bound state orbitals is important for obtaining electron
spectra. Indeed, it has been shown that with high-resolution
electron spectroscopy one can observe the energy splitting of the
molecular core-hole states 1$\mathrm{\sigma_g}$ and 1$\mathrm{\sigma_u}$ \cite{Ueda2009,Ueda2008,Schoffler2008}. In order to determine whether a pathway accesses a TSDCH molecular state or a SSDCH molecular state during the interaction of $\mathrm{N_{2}}$ with an FEL pulse, it is necessary at each time step of the propagation to project the delocalized inner-shell molecular orbitals onto inner-shell orbitals localized on atomic sites. We denote the DCH molecular states that involve the inner-shell molecular orbitals $1\sigma_g$ and $1\sigma_u$ by $|1\sigma_{g/u}1\sigma_{g/u}\rangle$. 

The delocalized molecular orbitals are expressed in terms of orbitals localized on atomic sites by $|1\sigma_g\rangle=\tfrac{1}{\sqrt{2}}(|1s_a\rangle+|1s_b\rangle)$ and $|1\sigma_u\rangle=\tfrac{1}{\sqrt{2}}(|1s_a\rangle-|1s_b\rangle)$ where $|1s_a\rangle$ and $|1s_b\rangle$ are $|1s\rangle$ orbitals localized on the atomic sites $a$ and $b$ of $\mathrm{N_{2}}$, respectively. At every time step, we check whether a DCH molecular state $|1\sigma_g1\sigma_g\rangle$, $|1\sigma_g1\sigma_u\rangle$ or $|1\sigma_u1\sigma_u\rangle$ has been accessed. Expressing the DCH molecular state $|1\sigma_g1\sigma_g\rangle$ in terms of orbitals localized on atomic sites, we obtain the following:

\begin{equation}
\begin{aligned}
|1\sigma_g1\sigma_g\rangle=\frac{1}{2}\left[|1s_a\rangle|1s_a\rangle+|1s_a\rangle|1s_b\rangle+|1s_b\rangle|1s_a\rangle+|1s_b\rangle|1s_b\rangle
\right],
\end{aligned}
\end{equation}

\begin{equation}
\begin{aligned}
|\langle 1s_a1s_a|1\sigma_g1\sigma_g\rangle|^2=|\langle 1s_a1s_b|1\sigma_g1\sigma_g\rangle|^2&\\
=|\langle 1s_b1s_a|1\sigma_g1\sigma_g\rangle|^2=|\langle 1s_b1s_b|1\sigma_g1\sigma_g\rangle|^2&=\frac{1}{4}.
\end{aligned}
\end{equation}
Thus, the formation of a DCH molecular state $|1\sigma_g1\sigma_g\rangle$ corresponds to 50\% probability of accessing a SSDCH molecular state, i.e. the $|1s_a1s_a\rangle$ or the $|1s_b1s_b\rangle$ state, and 50\% probability of accessing a TSDCH molecular state, i.e. the $|1s_a1s_b\rangle$ or the $|1s_b1s_a\rangle$ state. One can show that the same probabilities are obtained for the DCH molecular state $|1\sigma_u1\sigma_u\rangle$. 

Next, we show that the DCH molecular state $|1\sigma_g1\sigma_u\rangle$ corresponds to different probabilities of accessing a SSDCH molecular state versus a TSDCH molecular state depending on the spin of the state. Denoting by $S$ and $T$ the spatial part of a singlet or triplet DCH molecular state, respectively, we obtain
\begin{equation}
|1\sigma_g1\sigma_u\rangle^{S/T}=\frac{1}{\sqrt{2}}\left(|1\sigma_g\rangle|1\sigma_u\rangle\pm|1\sigma_u\rangle|1\sigma_g\rangle\right).
\end{equation}
Expressing the spatial part of the DCH molecular state $|1\sigma_g1\sigma_u\rangle^{T}$ in terms of orbitals localized on the atomic sites, we obtain 

\begin{equation}
\begin{aligned}
|1\sigma_g1\sigma_u\rangle^{T}=\frac{1}{\sqrt{2}}\left(|1s_b\rangle|1s_a\rangle-|1s_a\rangle|1s_b\rangle\right),
\end{aligned}
\end{equation}

\begin{equation}
\begin{aligned}
|\langle 1s_a1s_b|^{T}|1\sigma_g1\sigma_u\rangle^{T}|^2=1.
\end{aligned}
\end{equation}
and, thus, this state corresponds to 100\% probability of occupying a TSDCH molecular state. Similarly, expressing the DCH molecular state $|1\sigma_g1\sigma_u\rangle^{S}$ in terms of orbitals localized on the atomic sites, we obtain:

\begin{equation}
\begin{aligned}
|1\sigma_g1\sigma_u\rangle^{S}=\frac{1}{\sqrt{2}}\left(|1s_a\rangle|1s_a\rangle-|1s_b\rangle|1s_b\rangle
\right),
\end{aligned}
\end{equation}

\begin{equation}
\begin{aligned}
|\langle 1s_a1s_a|1\sigma_g1\sigma_u\rangle^{S}|^2=|\langle 1s_b1s_b|1\sigma_g1\sigma_u\rangle^{S}|^2&=\frac{1}{2}.
\end{aligned}
\end{equation}
and, thus, this state corresponds to 100\% probability of occupying a SSDCH molecular state. Taking also into account that a DCH molecular state has 75\% probability to be in the $|1\sigma_g1\sigma_u\rangle^{T}$ state and 25\% to be in the $|1\sigma_g1\sigma_u\rangle^{S}$ state it follows that the DCH state $|1\sigma_g1\sigma_u\rangle$ has 75\% probability to access a TSDCH molecular state and 25\% probability to access a SSDCH molecular state. The above probabilities are incorporated at every time step of our computations in order to calculate the probability of pathways that access TSDCH and SSDCH molecular states during the interaction of $\mathrm{N_{2}}$ with an FEL laser pulse. 

\section{Results}

\subsection{Electron Spectra}

\begin{figure}[h]
\centering
\includegraphics[width=\linewidth]{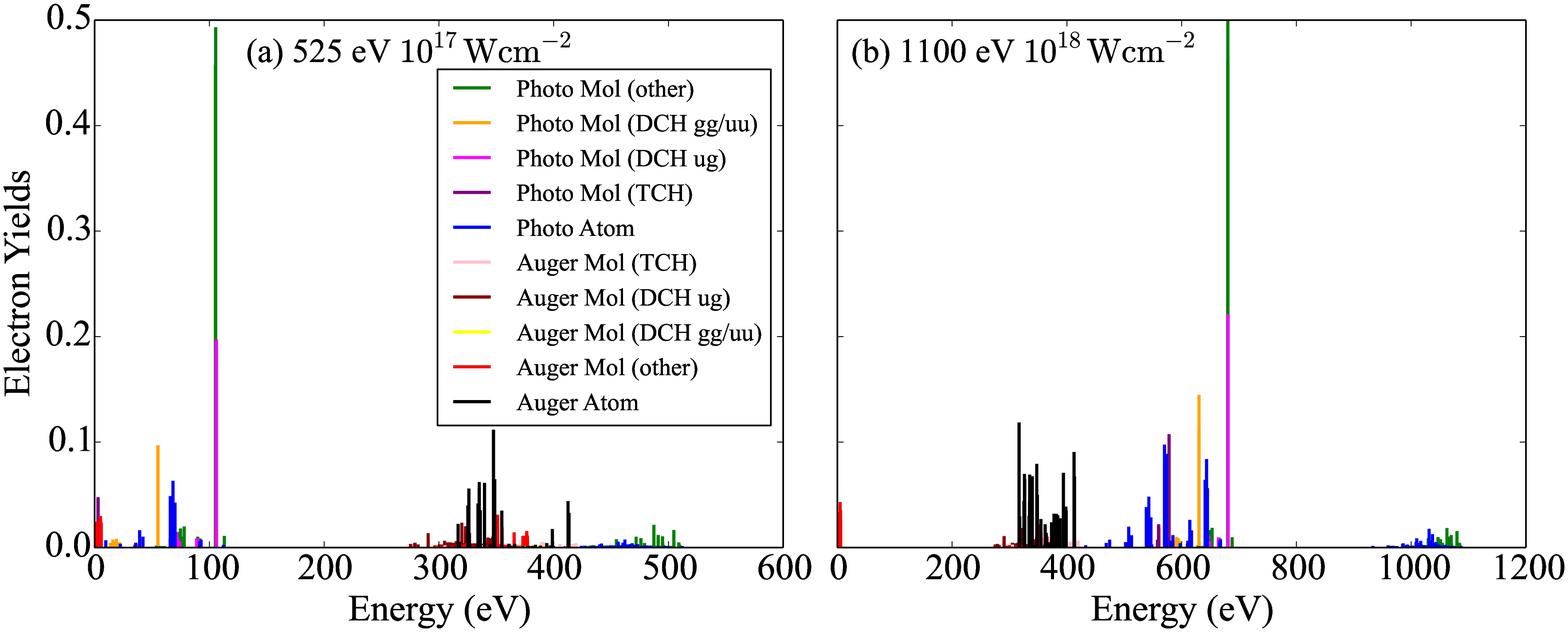}
\caption{Electron spectra resulting from the interaction of $\mathrm{N_{2}}$ with a 4 fs FWHM FEL pulse (a) with 525 eV photon energy and peak intensity of $\mathrm{10^{17}\;Wcm^{-2}}$ and (b) with 1100 eV photon energy and peak intensity of $\mathrm{10^{18}\;Wcm^{-2}}$. Transitions to or from a $|1\sigma_g1\sigma_u\rangle$ state are denoted by $\mathrm{DCH\,ug}$ while transitions involving a $|1\sigma_g1\sigma_g\rangle$ or $|1\sigma_u1\sigma_u\rangle$ state are denoted by $\mathrm{DCH\,gg/uu}$. } 
\label{fig:ES}
\end{figure}

Using the molecular and atomic Auger and photo-ionization yields, we plot in \fig{fig:ES} (a)((b)) the electron spectra generated by an FEL pulse with 525 eV (1100 eV) photon energy, with a full width at half maximum (FWHM) pulse duration of 4 fs and with a peak intensity of $\mathrm{10^{17}\;Wcm^{-2}}$ ($\mathrm{10^{18}\;Wcm^{-2}}$). We find that for both FEL pulses the transitions that involve a $|1\sigma_g1\sigma_u\rangle$ state or a TCH molecular state can not be easily discerned. The reason for this, is that these latter transitions have very small probability or that their energy is very close to the energies corresponding to other transitions.  In contrast, we find that the transition involving a $|1\sigma_g1\sigma_g\rangle$ or a $|1\sigma_u1\sigma_u\rangle$ state  can be easily discerned at 55 eV energy for the 525 eV  FEL pulse and at 630 eV for the 1100 eV FEL pulse. However, the transition involving a $|1\sigma_g1\sigma_g\rangle$ or a $|1\sigma_u1\sigma_u\rangle$ state  has 50\% probability to involve a TSDCH molecular and a SSDCH molecular state.
 Therefore, in the current formulation, we can not determine from the electron spectra whether a TSDCH and/or a SSDCH molecular state is formed.

\subsection{Comparison of atomic ion yields with experimental and theoretical results for different pulse durations}

\begin{figure*}[h]
\centering
\includegraphics[width=\linewidth]{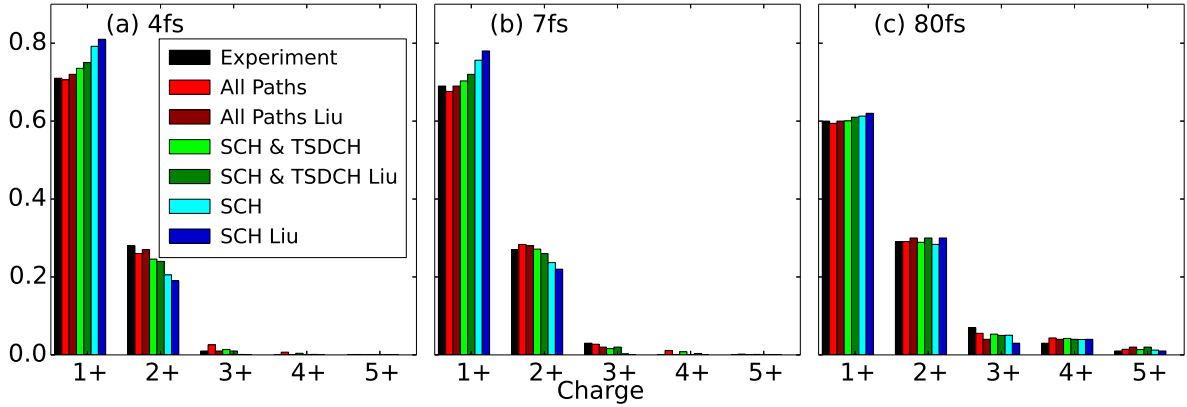}
\caption{Atomic ion yields for FEL pulses with (a) a total energy of 0.15 mJ, 77\% loss and 4 fs FWHM duration and b) a pulse energy of 0.26 mJ, 84\% loss and 7 fs FWHM duration c) a pulse energy of 0.26 mJ, 70\% loss, and 80 fs FWHM duration. Our results are compared with the experimental results in Ref. \cite{Hoener2010,Liu2016}. Atomic ion yields are obtained with all pathways accounted for as well as with certain pathways excluded and are compared with other theoretical results \cite{Liu2016}. }
\label{fig:IYcompFull}
\end{figure*}

In \fig{fig:IYcompFull}, we compare our results for the final atomic ion yields when $\mathrm{N_{2}}$ interacts with an FEL pulse with computations that use atomic instead of molecular orbitals \cite{Liu2016} and with experimental results \cite{Hoener2010}. To compare with experiment, we take the FEL flux to be given by

\begin{flalign}
\mathrm{J(x,y,t) = \rho(x,y)\Gamma_{ph}(t)},
\label{Jt1}
\end{flalign} 
where the transverse beam profile is given by 
 \begin{flalign}
\mathrm{\rho(x,y) = \frac{4ln2}{\pi\rho_x\rho_y}e^{-4ln2[(\tfrac{x}{\rho_x})^2+(\tfrac{y}{\rho_y})^2]}}
\label{Jt2}
\end{flalign}
with $\mathrm{\rho_x=2.2}$ $\mathrm{\mu}$m and $\mathrm{\rho_y=1.2}$ $\mathrm{\mu}$m in accord with the experimental parameters in ref. \cite{Hoener2010}.
The temporal profile for the rate of photons is given by 
 \begin{flalign}
\mathrm{\Gamma_{ph}(t) = \Gamma_{ph,0}e^{-4ln2(\frac{t}{\tau})^2}},
\label{Jt3}
\end{flalign}
where $\mathrm{\tau}$ is the FWHM duration of the pulse and $\mathrm{\Gamma_{ph,0}}$ is the maximum rate of photons 
 \begin{flalign}
\mathrm{\Gamma_{ph,0} = 2\sqrt{\frac{ln2}{\pi}}\frac{n_{ph}}{\tau}}.
\label{Jt4}
\end{flalign}
$\mathrm{n_{ph}=\frac{E_P}{\omega}}$ is the number of photons, each with energy $\omega$, in a pulse with energy $\mathrm{E_P}$. However, due to beam transport losses, a large portion of the energy does not reach the sample. As a result, we must use an effective energy, which is the total energy minus this loss.

We obtain the final atomic ion yields by computing the photon flux at different grid points (x,y), solving the rate equations  at each grid point and finally integrating over an area  of $\mathrm{10\mu m\times10\mu m}$. We  normalize so that the yields of the final atomic ions sum up to 1. When all pathways are accounted for, our results, displayed in \fig{fig:IYcompFull} (red color) compare very well with the experimental results \cite{Hoener2010} (black color). We also compute the atomic ion yields when we account only for pathways that go through SCH molecular states or through SCH plus TSDCH molecular states. We find that in all these cases our results agree well with the theoretical results in Ref.\cite{Liu2016}. As expected, the more pathways we exclude, the higher the yield is of the lower charged atomic ion. Indeed, when pathways that access solely SCH molecular states are accounted for, we find that the atomic ion yield for the $\mathrm{N^{+}}$ state is the highest. Moreover, as expected, we find that for longer FEL pulse durations the yields for higher charged atomic ions have larger values, since the number of inner-shell electrons ionizing by single-photon absorption increases.

\subsection{Contribution of SSDCH,TSDCH and TCH molecular states in atomic ion yields.}

\begin{figure}[h]
\centering
\includegraphics[width=0.95\linewidth]{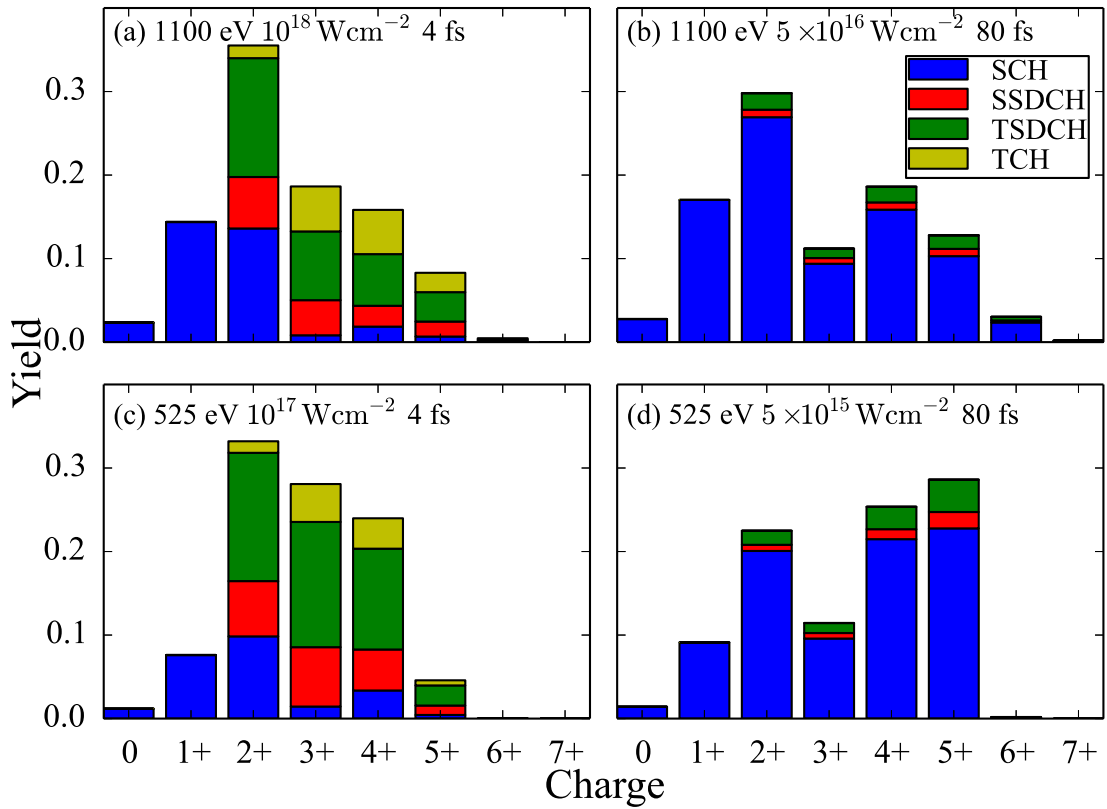}
\caption{Atomic ion yields for various FEL pulses. The contributions of pathways accessing SCH, TSDCH, SSDCH and TCH molecular states for each atomic ion yield.}
\label{fig:IonYields}
\end{figure}
 
In \fig{fig:IonYields}, for different FEL pulses, we compute the final atomic ion yields as well as the contribution of SCH, TSDCH, SSDCH and TCH molecular states to each of the final atomic ion yields. In \fig{fig:IonYields} (a) and (c), 
we find that for short duration of 4 fs FWHM and high intensity FEL pulses, 49\% and 56\% of all pathways contributing to all final atomic ion yields are pathways that have accessed TSDCH and TCH molecular states. Moreover, we find that the contribution of pathways that access TSDCH and TCH molecular states increases for higher charged atomic ion states. The reason for this, is that the contribution of pathways where two single-photon ionizations take place sequentially, i.e. before an Auger process takes place following the first single-photon ionization, is larger for atomic ions $\mathrm{N^{3+}}$ and $\mathrm{N^{4+}}$ compared to $\mathrm{N^{2+}}$. For instance, the contribution of pathways that have accessed SSDCH, TSDCH and TCH molecular states account for roughly 90\% of the $\mathrm{N^{5+}}$ yield for the short duration and intense FEL pulses, see \fig{fig:IonYields}(a) and (c). For the long duration of 80 fs FWHM and small intensity FEL pulses, we find that no more than 10\% of all pathways contributing to the atomic ion yields are pathways that have accessed SSDCH, TSDCH and TCH molecular states. In addition, the atomic ion yields of the higher charged states have much larger values for the 80 fs pulse rather than for the 4 fs FWHM FEL pulse. This is reasonable since for the long duration FEL pulse more single-photon ionization processes take place leading to the formation of higher charged atomic ions.

 
\subsection{Dependence of DCH and TCH molecular states on intensity.}
 
\begin{figure}[h]
\centering
\includegraphics[width=\linewidth]{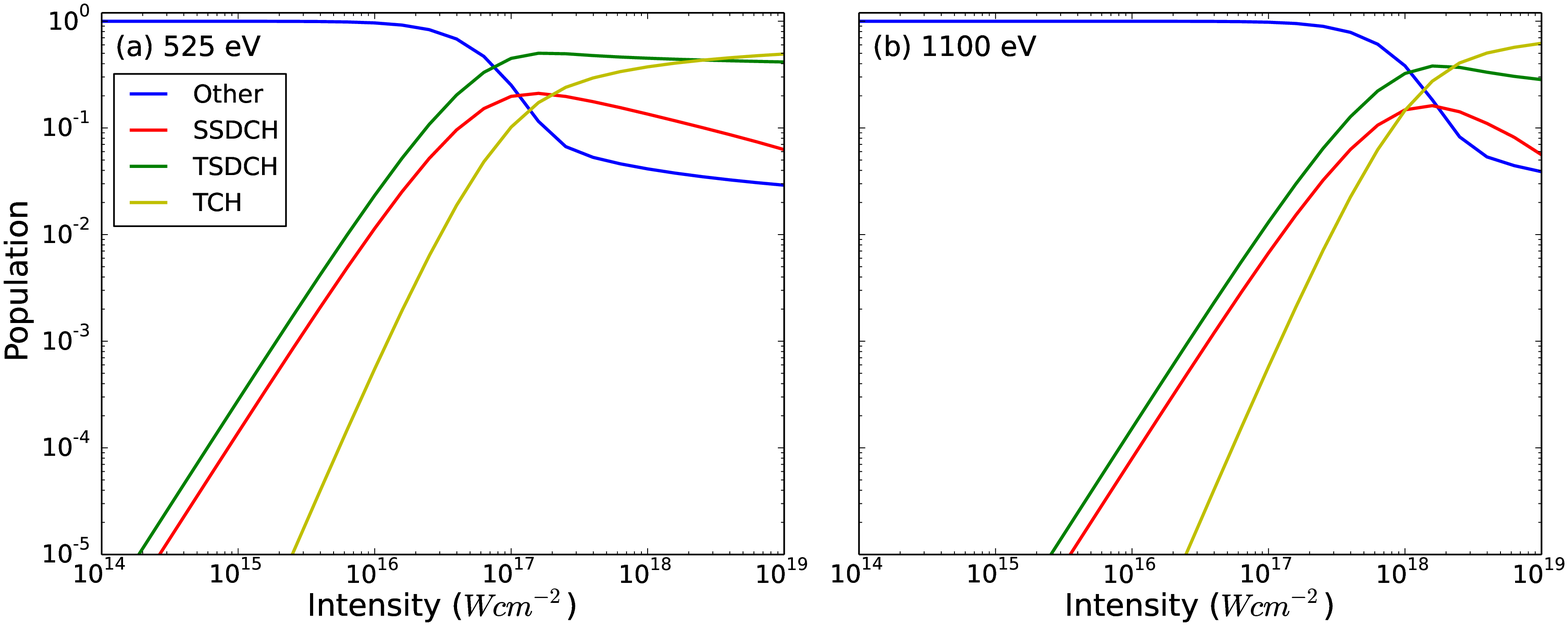}
\caption{Proportion of populations that access different core-hole states of $\mathrm{N_2}$ when driven  with a 4 fs FWHM and 525 eV (a) or 1100 eV (b) FEL pulse as a function of intensity.}
\label{fig:popTypes}
\end{figure}

In \fig{fig:popTypes}, we plot the population of pathways that accesses SCH, SSDCH, TSDCH and TCH molecular states as a function of intensity for a 4 fs FEL pulse for 525 eV photon energy (a) and for 1100 eV photon energy (b). 
We find that for the 525 eV (1100 eV) FEL pulse most of the population accesses multiple-core-hole molecular states for intensities above 10$^{17}$ Wcm$^{-2}$ (10$^{18}$ Wcm$^{-2}$). The intensity is higher for the higher photon energy pulse since the single-photon ionization cross sections are higher for the smaller photon energy FEL pulse. Moreover, we find that the contribution of TCH molecular states compared to TSDCH molecular states is higher for the 1100 eV rather than for the 525 eV FEL pulse for high intensities. The reason for this, is that more molecular states are energetically accessible with the 1100 eV photon energy FEL pulse.

\section{Conclusions}

In this work, we have computed the contribution to the final atomic ion yields of pathways that access molecular states with more than one inner-shell holes. We have identified the most efficient duration and intensity of a 525 eV and a 1100 eV  FEL pulse  in order to maximize the contribution of pathways that access DCH and TCH molecular states. Future work will take into consideration the motion of the nuclei which is essential in order to accurately describe the fragmentation process of FEL-driven molecules particularly for FEL pulses of long duration. The inclusion of spin in future work, could allow us to explicitly detect TSDCH or SSDCH states, based on section \ref{Projection}.

{\it Acknowledgments.} A.E. and H. I. B. Banks acknowledge the use of the Legion computational resources at UCL.

\bibliography{AugerNoURL} 

\providecommand{\newblock}{}
\begin{thebibliography}{10}
\expandafter\ifx\csname url\endcsname\relax
  \def\url#1{{\tt #1}}\fi
\expandafter\ifx\csname urlprefix\endcsname\relax\def\urlprefix{URL }\fi
\providecommand{\eprint}[2][]{\url{#2}}

\bibitem{FELhistory}
{Pellegrini} C 2012 {\em European Physical Journal H\/} {\bf 37} 659--708

\bibitem{Marangos2011}
Marangos J 2011 {\em Contemporary Physics\/} {\bf 52} 551--569

\bibitem{Ullrich2012}
Ullrich J, Rudenko A and Moshammer R 2012 {\em Annual Review of Physical
  Chemistry\/} {\bf 63} 635--660

\bibitem{Schlichting2012}
Schlichting I and Miao J 2012 {\em Current Opinion in Structural Biology\/}
  {\bf 22} 613 -- 626

\bibitem{Neutze}
Neutze R, Wouts R, van~der Spoel D, Weckert E and Hajdu J 2000 {\em Nature\/}
  {\bf 406}(6797) 752--757

\bibitem{Redecke227}
Redecke L, Nass K, DePonte D~P, White T~A, Rehders D, Barty A, Stellato F,
  Liang M, Barends T~R, Boutet S, Williams G~J, Messerschmidt M, Seibert M~M,
  Aquila A, Arnlund D, Bajt S, Barth T, Bogan M~J, Caleman C, Chao T~C, Doak
  R~B, Fleckenstein H, Frank M, Fromme R, Galli L, Grotjohann I, Hunter M~S,
  Johansson L~C, Kassemeyer S, Katona G, Kirian R~A, Koopmann R, Kupitz C, Lomb
  L, Martin A~V, Mogk S, Neutze R, Shoeman R~L, Steinbrener J, Timneanu N, Wang
  D, Weierstall U, Zatsepin N~A, Spence J~C~H, Fromme P, Schlichting I,
  Duszenko M, Betzel C and Chapman H~N 2013 {\em Science\/} {\bf 339} 227--230

\bibitem{Cederbaum1986}
Cederbaum L~S, Tarantelli F, Sgamellotti A and Schirmer J 1986 {\em The Journal
  of Chemical Physics\/} {\bf 85} 6513--6523

\bibitem{Tashiro2010}
Tashiro M, Ehara M, Fukuzawa H, Ueda K, Buth C, Kryzhevoi N~V and Cederbaum L~S
  2010 {\em The Journal of Chemical Physics\/} {\bf 132} 184302

\bibitem{Berrah2011}
Berrah N, Fang L, Murphy B, Osipov T, Ueda K, Kukk E, Feifel R, van~der Meulen
  P, Salen P, Schmidt H~T, Thomas R~D, Larsson M, Richter R, Prince K~C, Bozek
  J~D, Bostedt C, Wada S~i, Piancastelli M~N, Tashiro M and Ehara M 2011 {\em
  Proceedings of the National Academy of Sciences\/} {\bf 108} 16912--16915

\bibitem{Fang2010}
Fang L, Hoener M, Gessner O, Tarantelli F, Pratt S~T, Kornilov O, Buth C,
  G\"uhr M, Kanter E~P, Bostedt C, Bozek J~D, Bucksbaum P~H, Chen M, Coffee R,
  Cryan J, Glownia M, Kukk E, Leone S~R and Berrah N 2010 {\em Phys. Rev.
  Lett.\/} {\bf 105}(8) 083005

\bibitem{Fang2012}
Fang L, Osipov T, Murphy B, Tarantelli F, Kukk E, Cryan J~P, Glownia M,
  Bucksbaum P~H, Coffee R~N, Chen M, Buth C and Berrah N 2012 {\em Phys. Rev.
  Lett.\/} {\bf 109}(26) 263001

\bibitem{Salen2012}
Sal\'en P, van~der Meulen P, Schmidt H~T, Thomas R~D, Larsson M, Feifel R,
  Piancastelli M~N, Fang L, Murphy B, Osipov T, Berrah N, Kukk E, Ueda K, Bozek
  J~D, Bostedt C, Wada S, Richter R, Feyer V and Prince K~C 2012 {\em Phys.
  Rev. Lett.\/} {\bf 108}(15) 153003

\bibitem{Zhaunerchyk2015}
Zhaunerchyk V, Kami{\~{n}}"ska M, Mucke M, Squibb R~J, Eland J~H~D,
  Piancastelli M~N, Frasinski L~J, Grilj J, Koch M, McFarland B~K, Sistrunk E,
  G\"{u}hr M, Coffee R~N, Bostedt C, Bozek J~D, Sal\'en P, v~d Meulen P,
  Linusson P, Thomas R~D, Larsson M, Foucar L, Ullrich J, Motomura K, Mondal S,
  Ueda K, Richter R, Prince K~C, Takahashi O, Osipov T, Fang L, Murphy B~F,
  Berrah N and Feifel R 2015 {\em Journal of Physics B: Atomic, Molecular and
  Optical Physics\/} {\bf 48} 244003

\bibitem{Buth2012}
Buth C, Cryan J~P, Glownia J~M, Hoener M, Coffee R~N and Berrah N 2012 {\em The
  Journal of Chemical Physics\/} {\bf 136} 214310

\bibitem{Liu2016}
Liu J~C, Berrah N, Cederbaum L~S, Cryan J~P, Glownia J~M, Schafer K~J and Buth
  C 2016 {\em Journal of Physics B: Atomic, Molecular and Optical Physics\/}
  {\bf 49} 075602

\bibitem{Wallis2015}
Wallis A~O~G, Banks H~I~B and Emmanouilidou A 2015 {\em Phys. Rev. A\/} {\bf
  91}(6) 063402

\bibitem{Carravetta2013}
Carravetta V and {\AA}gren H 2013 {\em The Journal of Physical Chemistry A\/}
  {\bf 117} 6798--6802

\bibitem{Madsen2001}
Madsen L~B and Lambropoulos P 2001 {\em Journal of Physics B: Atomic, Molecular
  and Optical Physics\/} {\bf 34} 1855

\bibitem{Inhester2016}
Inhester L, Hanasaki K, Hao Y, Son S~K and Santra R 2016 {\em Phys. Rev. A\/}
  {\bf 94}(2) 023422

\bibitem{Molpro}
Werner H~J, Knowles P~J, Lindh R, Manby F~R, {Sch\"{u}tz} M {\em et~al.\/} 2010
  {MOLPRO}, a package of ab initio programs

\bibitem{Demekhin2011}
Demekhin P~V, Ehresmann A and Sukhorukov V~L 2011 {\em The Journal of Chemical
  Physics\/} {\bf 134} 024113

\bibitem{Herman1963}
Herman F and Skillman S 1963 {\em Atomic structure calculations\/}
  (Prentice-Hall, New Jersey)

\bibitem{Banks2017}
Banks H~I~B, Little D~A, Tennyson J and Emmanouilidou A 2017 {\em Phys. Chem.
  Chem. Phys.\/} {\bf 19}(30) 19794--19806

\bibitem{Beylerian2004}
Beylerian C and Cornaggia C 2004 {\em Journal of Physics B: Atomic, Molecular
  and Optical Physics\/} {\bf 37} L259

\bibitem{Ueda2009}
Ueda K, P{\"u}ttner R, Cherepkov N~A, Gel'mukhanov F and Ehara M 2009 {\em The
  European Physical Journal Special Topics\/} {\bf 169} 95--107

\bibitem{Ueda2008}
Ueda K 2008 {\em Science\/} {\bf 320} 884--885

\bibitem{Schoffler2008}
Sch{\"o}ffler M~S, Titze J, Petridis N, Jahnke T, Cole K, Schmidt L~P~H, Czasch
  A, Akoury D, Jagutzki O, Williams J~B, Cherepkov N~A, Semenov S~K, McCurdy
  C~W, Rescigno T~N, Cocke C~L, Osipov T, Lee S, Prior M~H, Belkacem A, Landers
  A~L, Schmidt-B{\"o}cking H, Weber T and D{\"o}rner R 2008 {\em Science\/}
  {\bf 320} 920--923

\bibitem{Hoener2010}
Hoener M, Fang L, Kornilov O, Gessner O, Pratt S~T, G\"uhr M, Kanter E~P, Blaga
  C, Bostedt C, Bozek J~D, Bucksbaum P~H, Buth C, Chen M, Coffee R, Cryan J,
  DiMauro L, Glownia M, Hosler E, Kukk E, Leone S~R, McFarland B, Messerschmidt
  M, Murphy B, Petrovic V, Rolles D and Berrah N 2010 {\em Phys. Rev. Lett.\/}
  {\bf 104}(25) 253002

\end{thebibliography}

\end{document}